\documentstyle{IOP}
\def\temp{1.34}%
\let\tempp=\relax
\expandafter\ifx\csname psboxversion\endcsname\relax
\else
    \ifdim\temp cm>\psboxversion cm
    \else
      \let\temp=\psboxversion
      \let\tempp= 
    \fi
\fi
\tempp
\let\psboxversion=\temp
\catcode`\@=11
\def\psfortextures{
\def\PSspeci@l##1##2{%
\special{illustration ##1\space scaled ##2}%
}}%
\def\psfordvitops{
\def\PSspeci@l##1##2{%
\special{dvitops: import ##1\space \the\drawingwd \the\drawinght}%
}}%
\def\psfordvips{
\def\PSspeci@l##1##2{%
\d@my=0.1bp \d@mx=\drawingwd \divide\d@mx by\d@my
\includegraphics{##1\space}}}%
\def\psforoztex{
\def\PSspeci@l##1##2{%
\special{##1 \space
      ##2 1000 div dup scale
      \number-\psllx\space \number-\pslly\space translate
}}}%
\def\psfordvitps{
\def\psdimt@n@sp##1{\d@mx=##1\relax\edef\psn@sp{\number\d@mx}}
\def\PSspeci@l##1##2{%
\special{dvitps: Include0 "psfig.psr"}
\psdimt@n@sp{\drawingwd}
\special{dvitps: Literal "\psn@sp\space"}
\psdimt@n@sp{\drawinght}
\special{dvitps: Literal "\psn@sp\space"}
\psdimt@n@sp{\psllx bp}
\special{dvitps: Literal "\psn@sp\space"}
\psdimt@n@sp{\pslly bp}
\special{dvitps: Literal "\psn@sp\space"}
\psdimt@n@sp{\psurx bp}
\special{dvitps: Literal "\psn@sp\space"}
\psdimt@n@sp{\psury bp}
\special{dvitps: Literal "\psn@sp\space startTexFig\space"}
\special{dvitps: Include1 "##1"}
\special{dvitps: Literal "endTexFig\space"}
}}%
\def\psfordvialw{
\def\PSspeci@l##1##2{
\special{language "PostScript",
position = "bottom left",
literal "  \psllx\space \pslly\space translate
  ##2 1000 div dup scale
  -\psllx\space -\pslly\space translate",
include "##1"}
}}%
\def\psforptips{
\def\PSspeci@l##1##2{{
\d@mx=\psurx bp
\advance \d@mx by -\psllx bp
\divide \d@mx by 1000\multiply\d@mx by \xscale
\incm{\d@mx}
\let\tmpx\dimincm
\d@my=\psury bp
\advance \d@my by -\pslly bp
\divide \d@my by 1000\multiply\d@my by \xscale
\incm{\d@my}
\let\tmpy\dimincm
\d@mx=-\psllx bp
\divide \d@mx by 1000\multiply\d@mx by \xscale
\d@my=-\pslly bp
\divide \d@my by 1000\multiply\d@my by \xscale
\at(\d@mx;\d@my){\special{ps:##1 x=\tmpx, y=\tmpy}}
}}}%
\def\psonlyboxes{
\def\PSspeci@l##1##2{%
\at(0cm;0cm){\boxit{\vbox to\drawinght
  {\vss\hbox to\drawingwd{\at(0cm;0cm){\hbox{({\tt##1})}}\hss}}}}
}}%
\def\psloc@lerr#1{%
\let\savedPSspeci@l=\PSspeci@l%
\def\PSspeci@l##1##2{%
\at(0cm;0cm){\boxit{\vbox to\drawinght
  {\vss\hbox to\drawingwd{\at(0cm;0cm){\hbox{({\tt##1}) #1}}\hss}}}}
\let\PSspeci@l=\savedPSspeci@l
}}%
\newread\pst@mpin
\newdimen\drawinght\newdimen\drawingwd
\newdimen\psxoffset\newdimen\psyoffset
\newbox\drawingBox
\newcount\xscale \newcount\yscale \newdimen\pscm\pscm=1cm
\newdimen\d@mx \newdimen\d@my
\newdimen\pswdincr \newdimen\pshtincr
\let\ps@nnotation=\relax
{\catcode`\|=0 |catcode`|\=12 |catcode`|
|catcode`#=12 |catcode`*=14
|xdef|backslashother{\}*
|xdef|percentother{
|xdef|tildeother{~}*
|xdef|sharpother{#}*
}%
\def\R@moveMeaningHeader#1:->{}%
\def\uncatcode#1{%
\edef#1{\expandafter\R@moveMeaningHeader\meaning#1}}%
\def\execute#1{#1}
\def\psm@keother#1{\catcode`#112\relax}
\def\executeinspecs#1{%
\execute{\begingroup\let\do\psm@keother\dospecials\catcode`\^^M=9#1\endgroup}}%
\def\@mpty{}%
\def\matchexpin#1#2{
  \fi%
  \edef\tmpb{{#2}}%
  \expandafter\makem@tchtmp\tmpb%
  \edef\tmpa{#1}\edef\tmpb{#2}%
  \expandafter\expandafter\expandafter\m@tchtmp\expandafter\tmpa\tmpb\endm@tch%
  \if\match%
}%
\def\matchin#1#2{%
  \fi%
  \makem@tchtmp{#2}%
  \m@tchtmp#1#2\endm@tch%
  \if\match%
}%
\def\makem@tchtmp#1{\def\m@tchtmp##1#1##2\endm@tch{%
  \def\tmpa{##1}\def\tmpb{##2}\let\m@tchtmp=\relax%
  \ifx\tmpb\@mpty\def\match{YN}%
  \else\def\match{YY}\fi%
}}%
\def\incm#1{{\psxoffset=1cm\d@my=#1
 \d@mx=\d@my
  \divide\d@mx by \psxoffset
  \xdef\dimincm{\number\d@mx.}
  \advance\d@my by -\number\d@mx cm
  \multiply\d@my by 100
 \d@mx=\d@my
  \divide\d@mx by \psxoffset
  \edef\dimincm{\dimincm\number\d@mx}
  \advance\d@my by -\number\d@mx cm
  \multiply\d@my by 100
 \d@mx=\d@my
  \divide\d@mx by \psxoffset
  \xdef\dimincm{\dimincm\number\d@mx}
}}%
\newif\ifNotB@undingBox
\newhelp\PShelp{Proceed: you'll have a 5cm square blank box instead of
your graphics (Jean Orloff).}%
\def\s@tsize#1 #2 #3 #4\@ndsize{
  \def\psllx{#1}\def\pslly{#2}%
  \def\psurx{#3}\def\psury{#4}
  \ifx\psurx\@mpty\NotB@undingBoxtrue
  \else
    \drawinght=#4bp\advance\drawinght by-#2bp
    \drawingwd=#3bp\advance\drawingwd by-#1bp
  \fi
  }%
\def\sc@nBBline#1:#2\@ndBBline{\edef\p@rameter{#1}\edef\v@lue{#2}}%
\def\g@bblefirstblank#1#2:{\ifx#1 \else#1\fi#2}%
{\catcode`\%=12
\xdef\B@undingBox{
\def\ReadPSize#1{
 \readfilename#1\relax
 \let\PSfilename=\lastreadfilename
 \openin\pst@mpin=#1\relax
 \ifeof\pst@mpin \errhelp=\PShelp
   \errmessage{I haven't found your postscript file (\PSfilename)}%
   \psloc@lerr{was not found}%
   \s@tsize 0 0 142 142\@ndsize
   \closein\pst@mpin
 \else
   \if\matchexpin{\GlobalInputList}{, \lastreadfilename}%
   \else\xdef\GlobalInputList{\GlobalInputList, \lastreadfilename}%
     \immediate\write\psbj@inaux{\lastreadfilename,}%
   \fi%
   \loop
     \executeinspecs{\catcode`\ =10\global\read\pst@mpin to\n@xtline}%
     \ifeof\pst@mpin
       \errhelp=\PShelp
       \errmessage{(\PSfilename) is not an Encapsulated PostScript File:
           I could not find any \B@undingBox: line.}%
       \edef\v@lue{0 0 142 142:}%
       \psloc@lerr{is not an EPSFile}%
       \NotB@undingBoxfalse
     \else
       \expandafter\sc@nBBline\n@xtline:\@ndBBline
       \ifx\p@rameter\B@undingBox\NotB@undingBoxfalse
         \edef\t@mp{%
           \expandafter\g@bblefirstblank\v@lue\space\space\space}%
         \expandafter\s@tsize\t@mp\@ndsize
       \else\NotB@undingBoxtrue
       \fi
     \fi
   \ifNotB@undingBox\repeat
   \closein\pst@mpin
 \fi
\message{#1}%
}%
\def\psboxto(#1;#2)#3{\vbox{%
   \ReadPSize{#3}%
   \advance\pswdincr by \drawingwd
   \advance\pshtincr by \drawinght
   \divide\pswdincr by 1000
   \divide\pshtincr by 1000
   \d@mx=#1
   \ifdim\d@mx=0pt\xscale=1000
         \else \xscale=\d@mx \divide \xscale by \pswdincr\fi
   \d@my=#2
   \ifdim\d@my=0pt\yscale=1000
         \else \yscale=\d@my \divide \yscale by \pshtincr\fi
   \ifnum\yscale=1000
         \else\ifnum\xscale=1000\xscale=\yscale
                    \else\ifnum\yscale<\xscale\xscale=\yscale\fi
              \fi
   \fi
   \divide\drawingwd by1000 \multiply\drawingwd by\xscale
   \divide\drawinght by1000 \multiply\drawinght by\xscale
   \divide\psxoffset by1000 \multiply\psxoffset by\xscale
   \divide\psyoffset by1000 \multiply\psyoffset by\xscale
   \global\divide\pscm by 1000
   \global\multiply\pscm by\xscale
   \multiply\pswdincr by\xscale \multiply\pshtincr by\xscale
   \ifdim\d@mx=0pt\d@mx=\pswdincr\fi
   \ifdim\d@my=0pt\d@my=\pshtincr\fi
   \message{scaled \the\xscale}%
 \hbox to\d@mx{\hss\vbox to\d@my{\vss
   \global\setbox\drawingBox=\hbox to 0pt{\kern\psxoffset\vbox to 0pt{%
      \kern-\psyoffset
      \PSspeci@l{\PSfilename}{\the\xscale}%
      \vss}\hss\ps@nnotation}%
   \global\wd\drawingBox=\the\pswdincr
   \global\ht\drawingBox=\the\pshtincr
   \global\drawingwd=\pswdincr
   \global\drawinght=\pshtincr
   \baselineskip=0pt
   \copy\drawingBox
 \vss}\hss}%
  \global\psxoffset=0pt
  \global\psyoffset=0pt
  \global\pswdincr=0pt
  \global\pshtincr=0pt 
  \global\pscm=1cm 
}}%
\def\psboxscaled#1#2{\vbox{%
  \ReadPSize{#2}%
  \xscale=#1
  \message{scaled \the\xscale}%
  \divide\pswdincr by 1000 \multiply\pswdincr by \xscale
  \divide\pshtincr by 1000 \multiply\pshtincr by \xscale
  \divide\psxoffset by1000 \multiply\psxoffset by\xscale
  \divide\psyoffset by1000 \multiply\psyoffset by\xscale
  \divide\drawingwd by1000 \multiply\drawingwd by\xscale
  \divide\drawinght by1000 \multiply\drawinght by\xscale
  \global\divide\pscm by 1000
  \global\multiply\pscm by\xscale
  \global\setbox\drawingBox=\hbox to 0pt{\kern\psxoffset\vbox to 0pt{%
     \kern-\psyoffset
     \PSspeci@l{\PSfilename}{\the\xscale}%
     \vss}\hss\ps@nnotation}%
  \advance\pswdincr by \drawingwd
  \advance\pshtincr by \drawinght
  \global\wd\drawingBox=\the\pswdincr
  \global\ht\drawingBox=\the\pshtincr
  \global\drawingwd=\pswdincr
  \global\drawinght=\pshtincr
  \baselineskip=0pt
  \copy\drawingBox
  \global\psxoffset=0pt
  \global\psyoffset=0pt
  \global\pswdincr=0pt
  \global\pshtincr=0pt 
  \global\pscm=1cm
}}%
\def\psbox#1{\psboxscaled{1000}{#1}}%
\newif\ifn@teof\n@teoftrue
\newif\ifc@ntrolline
\newif\ifmatch
\newread\j@insplitin
\newwrite\j@insplitout
\newwrite\psbj@inaux
\immediate\openout\psbj@inaux=psbjoin.aux
\immediate\write\psbj@inaux{\string\joinfiles}%
\immediate\write\psbj@inaux{\jobname,}%
\def\toother#1{\ifcat\relax#1\else\expandafter%
  \toother@ux\meaning#1\endtoother@ux\fi}%
\def\toother@ux#1 #2#3\endtoother@ux{\def\tmp{#3}%
  \ifx\tmp\@mpty\def\tmp{#2}\let\next=\relax%
  \else\def\next{\toother@ux#2#3\endtoother@ux}\fi%
\next}%
\let\readfilenamehook=\relax
\def\re@d{\expandafter\re@daux}
\def\re@daux{\futurelet\nextchar\stopre@dtest}%
\def\re@dnext{\xdef\lastreadfilename{\lastreadfilename\nextchar}%
  \afterassignment\re@d\let\nextchar}%
\def\stopre@d{\egroup\readfilenamehook}%
\def\stopre@dtest{%
  \ifcat\nextchar\relax\let\nextread\stopre@d
  \else
    \ifcat\nextchar\space\def\nextread{%
      \afterassignment\stopre@d\chardef\nextchar=`}%
    \else\let\nextread=\re@dnext
      \toother\nextchar
      \edef\nextchar{\tmp}%
    \fi
  \fi\nextread}%
\def\readfilename{\bgroup%
  \let\\=\backslashother \let\%=\percentother \let\~=\tildeother
  \let\#=\sharpother \xdef\lastreadfilename{}%
  \re@d}%
\xdef\GlobalInputList{\jobname}%
\def\psnewinput{%
  \def\readfilenamehook{
    \if\matchexpin{\GlobalInputList}{, \lastreadfilename}%
    \else\xdef\GlobalInputList{\GlobalInputList, \lastreadfilename}%
      \immediate\write\psbj@inaux{\lastreadfilename,}%
    \fi%
    \ps@ldinput\lastreadfilename\relax%
    \let\readfilenamehook=\relax%
  }\readfilename%
}%
\expandafter\ifx\csname @@input\endcsname\relax    
  \immediate\let\ps@ldinput=\input\def\input{\psnewinput}%
\else
  \immediate\let\ps@ldinput=\@@input
  \def\@@input{\psnewinput}%
\fi%
\def\nowarnopenout{%
 \def\warnopenout##1##2{%
   \readfilename##2\relax
   \message{\lastreadfilename}%
   \immediate\openout##1=\lastreadfilename\relax}}%
\def\warnopenout#1#2{%
 \readfilename#2\relax
 \def\t@mp{TrashMe,psbjoin.aux,psbjoint.tex,}\uncatcode\t@mp
 \if\matchexpin{\t@mp}{\lastreadfilename,}%
 \else
   \immediate\openin\pst@mpin=\lastreadfilename\relax
   \ifeof\pst@mpin
     \else
     \errhelp{If the content of this file is so precious to you, abort (ie
press x or e) and rename it before retrying.}%
     \errmessage{I'm just about to replace your file named \lastreadfilename}%
   \fi
   \immediate\closein\pst@mpin
 \fi
 \message{\lastreadfilename}%
 \immediate\openout#1=\lastreadfilename\relax}%
{\catcode`\%=12\catcode`\*=14
\gdef\splitfile#1{*
 \readfilename#1\relax
 \immediate\openin\j@insplitin=\lastreadfilename\relax
 \ifeof\j@insplitin
   \message{! I couldn't find and split \lastreadfilename!}*
 \else
   \immediate\openout\j@insplitout=TrashMe
   \message{< Splitting \lastreadfilename\space into}*
   \loop
     \ifeof\j@insplitin
       \immediate\closein\j@insplitin\n@teoffalse
     \else
       \n@teoftrue
       \executeinspecs{\global\read\j@insplitin to\spl@tinline\expandafter
         \ch@ckbeginnewfile\spl@tinline
       \ifc@ntrolline
       \else
         \toks0=\expandafter{\spl@tinline}*
         \immediate\write\j@insplitout{\the\toks0}*
       \fi
     \fi
   \ifn@teof\repeat
   \immediate\closeout\j@insplitout
 \fi\message{>}*
}*
\gdef\ch@ckbeginnewfile#1
 \def\t@mp{#1}*
 \ifx\@mpty\t@mp
   \def\t@mp{#3}*
   \ifx\@mpty\t@mp
     \global\c@ntrollinefalse
   \else
     \immediate\closeout\j@insplitout
     \warnopenout\j@insplitout{#2}*
     \global\c@ntrollinetrue
   \fi
 \else
   \global\c@ntrollinefalse
 \fi}*
\gdef\joinfiles#1\into#2{*
 \message{< Joining following files into}*
 \warnopenout\j@insplitout{#2}*
 \message{:}*
 {*
 \edef\w@##1{\immediate\write\j@insplitout{##1}}*
\w@{
\w@{
\w@{
\w@{
\w@{
\w@{
\w@{
\w@{
\w@{
\w@{
\w@{\string\input\space psbox.tex}*
\w@{\string\splitfile{\string\jobname}}*
\w@{\string\let\string\autojoin=\string\relax}*
}*
 \expandafter\tre@tfilelist#1, \endtre@t
 \immediate\closeout\j@insplitout
 \message{>}*
}*
\gdef\tre@tfilelist#1, #2\endtre@t{*
 \readfilename#1\relax
 \ifx\@mpty\lastreadfilename
 \else
   \immediate\openin\j@insplitin=\lastreadfilename\relax
   \ifeof\j@insplitin
     \errmessage{I couldn't find file \lastreadfilename}*
   \else
     \message{\lastreadfilename}*
     \immediate\write\j@insplitout{
     \executeinspecs{\global\read\j@insplitin to\oldj@ininline}*
     \loop
       \ifeof\j@insplitin\immediate\closein\j@insplitin\n@teoffalse
       \else\n@teoftrue
         \executeinspecs{\global\read\j@insplitin to\j@ininline}*
         \toks0=\expandafter{\oldj@ininline}*
         \let\oldj@ininline=\j@ininline
         \immediate\write\j@insplitout{\the\toks0}*
       \fi
     \ifn@teof
     \repeat
   \immediate\closein\j@insplitin
   \fi
   \tre@tfilelist#2, \endtre@t
 \fi}*
}%
\def\autojoin{%
 \immediate\write\psbj@inaux{\string\into{psbjoint.tex}}%
 \immediate\closeout\psbj@inaux
 \expandafter\joinfiles\GlobalInputList\into{psbjoint.tex}%
}%
\def\centinsert#1{\midinsert\line{\hss#1\hss}\endinsert}%
\def\psannotate#1#2{\vbox{%
  \def\ps@nnotation{#2\global\let\ps@nnotation=\relax}#1}}%
\def\pscaption#1#2{\vbox{%
   \setbox\drawingBox=#1
   \copy\drawingBox
   \vskip\baselineskip
   \vbox{\hsize=\wd\drawingBox\setbox0=\hbox{#2}%
     \ifdim\wd0>\hsize
       \noindent\unhbox0\tolerance=5000
    \else\centerline{\box0}%
    \fi
}}}%
\def\at(#1;#2)#3{\setbox0=\hbox{#3}\ht0=0pt\dp0=0pt
  \rlap{\kern#1\vbox to0pt{\kern-#2\box0\vss}}}%
\newdimen\gridht \newdimen\gridwd
\def\gridfill(#1;#2){%
  \setbox0=\hbox to 1\pscm
  {\vrule height1\pscm width.4pt\leaders\hrule\hfill}%
  \gridht=#1
  \divide\gridht by \ht0
  \multiply\gridht by \ht0
  \gridwd=#2
  \divide\gridwd by \wd0
  \multiply\gridwd by \wd0
  \advance \gridwd by \wd0
  \vbox to \gridht{\leaders\hbox to\gridwd{\leaders\box0\hfill}\vfill}}%
\def\fillinggrid{\at(0cm;0cm){\vbox{%
  \gridfill(\drawinght;\drawingwd)}}}%
\def\textleftof#1:{%
  \setbox1=#1
  \setbox0=\vbox\bgroup
    \advance\hsize by -\wd1 \advance\hsize by -2em}%
\def\textrightof#1:{%
  \setbox0=#1
  \setbox1=\vbox\bgroup
    \advance\hsize by -\wd0 \advance\hsize by -2em}%
\def\endtext{%
  \egroup
  \hbox to \hsize{\valign{\vfil##\vfil\cr%
\box0\cr%
\noalign{\hss}\box1\cr}}}%
\def\frameit#1#2#3{\hbox{\vrule width#1\vbox{%
  \hrule height#1\vskip#2\hbox{\hskip#2\vbox{#3}\hskip#2}%
        \vskip#2\hrule height#1}\vrule width#1}}%
\def\boxit#1{\frameit{0.4pt}{0pt}{#1}}%
\catcode`\@=12 
 \psfordvips   

\begin{document}
\title[Multi-photon Dissociation Thresholds of Nitrogen Oxide]{
Driven Morse Oscillator: \\
Model for Multi-photon Dissociation of  \\
Nitrogen Oxide}

\author{Julian Juhi-Lian Ting }

\affiliation{Low-Energy Physics Institute, Tsing-Hua University, \\
Hsin-Chu, Taiwan  30043, Republic of China \footnote{E-mail address:
jlting@phys.nthu.edu.tw}}

\date{\today }
\maketitle
\begin{abstract}

Within a one-dimensional semi-classical model with a Morse potential 
the possibility of infrared multi-photon dissociation of vibrationally excited
nitrogen oxide was studied.
The dissociation
thresholds of typical driving forces and couplings 
were found to be similar, which indicates that the results were robust 
to variations of the potential and of the definition of dissociation rate.

\end{abstract}
\section{Introduction}

Experimental microwave multi-photon dissociation of nitrogen oxide (NO) is not yet
possible, whereas the corresponding optical photodissociation is readily
observed via electronic transitions. 
However, it is important to know the possibility
and how one can make it, firstly because NO is abundant in nature 
and in living
bodies, and secondly because the intensity of lasers has 
reached the order of magnitude needed to do
the experiment.
Although hydrogen fluoride (HF) has a smaller dissociation energy,
from an experimental view, because of the stability, simplicity and
well documented properties of NO, 
easier preparation of intense molecular beams of NO than of HF, 
and more efficient probing,
NO has been chosen for investigation instead of HF.
Although experiments were formerly conducted as NO mostly in the ground state
prepared by supersonic expansion, it has become possible to
transfer population efficiently to other states:
\cite{YKW} report experiments in which NO
molecules were populated into an initial vibrational state as great as $
n = 25 $
by stimulated-emission pumping (SEP) whereas \cite{SKSB} showed recently
how to populate to a state $n = 6$ by a stimulated
Raman process involving adiabatic passage (STIRAP) method using pulsed
lasers.
In our  study we chose parameters for NO whenever molecular 
properties were needed for our calculations, so that future experiments can 
be easily compared with the results presented here.
The behaviour of NO is expected to be typical of that of other
diatomic molecules.

Theoretically much previous work concentrated on the corresponding atomic systems.
Similar to the present work \cite{BS1} did a calculation 
on the hydrogen
atom and \cite{BFV} made a calculation on HF.  
But no previous calculation of
the dissociation thresholds of every initial eigenstates has been performed.
In this paper the Morse potential-energy function for the 
inter-atomic interaction was used; only the 
vibrational levels were considered presently because
\cite{M} and \cite{P} have pointed out that 
the rotational influence on vibration can be taken
into account by suitably adjusting the Morse parameters. 

The Morse potential was used by
\cite{LJS1,LJS2}
to investigate the interaction of
atoms and molecules with solid surfaces in the early days.
\cite{IR,WP} 
studied the driven Morse oscillator as a model for infrared 
multi-photon 
excitation and for dissociation of
molecules, stimulated by the possibility of laser isotope separation and 
bond-selective chemistry.
The Morse potential lacks the complication
of infinitely  many bound eigenstates and the singularity of the Coulomb
potential at the origin of the hydrogen atom.

The methods to calculate multi-photon dissociation  are
generally divided into nonlinear classical-mechanical
methods, like that of \cite{GM} and semi-quantum methods like that of \cite{HM2,HM1,TM2,TM1}.
The latter
approach is used in this study. 

\section{Model}

We consider an isolated non-rotating NO molecule interacting
with a plane-polarised harmonic laser field.
The dimensionless Hamiltonian of a free Morse oscillator for a diatomic molecule is
\begin{equation}
H_0={p^2 \over 2} + {(1-e^{- z})^2 \over 2},
\end{equation}
in which
$z=\alpha(r-r_e)$ denotes displacement of inter-atomic distance from
equilibrium,
$p=p_z/\sqrt{2mD}$, and $p_z$ is the  momentum conjugate to $z$.
The parameters of NO are given by \cite{HH} and summarised
in the TABLE.
The Hamiltonian for the Morse oscillator in the presence of a typical harmonic
laser field reads
\begin{equation}
H = H_0 -{{A \Omega} \over 2} \mu (z) \Phi (\Omega t),
\label{H}
\end{equation}
in which $\mu$ is the dipole moment operator.
In the formula above two dimensionless variables are introduced,
namely
$\Omega = \omega_L / \omega_0$ with
$\omega _L $ the laser frequency,
and the dimensionless field strength $A=q E_L/\alpha \Omega D$ with
$E_L$ the field strength of the laser and $q$ is the effective dipole charge
of the molecule.
In the numerical calculations below various types of couplings are
considered.

\begin{table}
\begin{center}
\begin{minipage}{6cm}
\begin{tabular}{ccc}
parameter & symbol & value\\
range parameter of the potential & $\alpha$ &$2.7675
\times {10^8} cm^{-1}$ \\
reduced mass &$m$ &  $7.4643u$ \\
dissociation energy &$D$ &  $6.4968 eV$ \\
Morse frequency &$\omega _0 = \sqrt{2 D {\alpha}^2/m}$
&       $1904.2 cm^{-1}$\\
\end{tabular}
\end{minipage}
\end{center}
\caption{Typical parameters for NO.}
\end{table}

The discrete eigenfunctions of the free Morse oscillator 
were given by \cite{M} 
%
and denoted by $| n >$ with the corresponding
dimensionless eigenvalues equal to
\begin{equation}
E_n={{(n+ {1 \over 2})} \over \lambda}-
{{(n+ {1\over 2})^2} \over {2 \lambda^2}},
\end{equation}
in which $\lambda = \sqrt {2 m D} / \alpha \hbar = 55.04 $.
The term multi-photon is generally applied to this system,
as each photon has energy
about $0.24 eV$ near the Morse frequency ($ \omega_0$), 
whereas the energy difference
between the ground state and the first excited state of the free Morse
oscillator is about $0.23 eV$; about 28 photons must be absorbed for the 
transition from the ground state to the continuum.

Our main  concern is the transition between each excited state
of the free Morse oscillator
and dissociation. 
With various initial states $| n >$,
driving frequencies and driving amplitudes the
dissociation periods required might change dramatically.
The problems of this calculation are whether our model can really represent the
molecular system for which we intend it for and what is our definition of
dissociation. How robust  are
the results?
To address these questions we present  two typical definitions of dissociation
and driving forces.

\section{Numerical results}
The following numerical results were obtained using a method
described and verified
previously by \cite{TYJ}, which is a fast Fourier-transformed grid method
also considered by 
\cite{F,ALL}.
\subsection{Dipolar coupling}

Following \cite{WP,GM} on  a driven Morse oscillator, as a
consistency check in the first instance, a
cosinusoidal wave form was used. Hence the Hamiltonian in Eq. (\ref{H})
reads
\begin{equation}
H = H_0 -{{A \Omega} \over 2}{z} \cos (\Omega t).
\label{H1}
\end{equation}
According to \cite{GM,CB} for quantum chaos
the definition of dissociation rate is
\begin{equation}
 P_{diss} (t) = 1 - \sum_n | < n  | \psi(t) > | ^2.
\label{P'}
\end{equation}
We {\it define} $P_{diss}(t)=0.05$ for the molecule to satisfy the dissociation
condition.
If the molecule cannot reach such a dissociation 
rate within 300 cycles we
consider it to be not dissociable under such conditions of field intensity and frequency.
This definition is almost equivalent to comparing the dissociation rate
after 300 cycles. According to \cite{ALL} the latter should be easier to calculate. However,
these two definitions are distinct.
Furthermore, the definition of $P_{diss}$ is only approximate because, when the field
is applied, the projection onto the bound states fails to
represent the true bound-state population unless one assumes that the
electro-magnetic field can be terminated instantaneously.
The values 300 and 0.05 are set according to previous experience,
although there are some guiding principles:
300 cycles corresponds to a pulse of duration about 5 picosecond, which is the time
scale for bound dissociation to occur.
These choices are analogous to those used in the calculation of 
atomic ionisation 
by
\cite{BS1}.

\begin{figure}
\begin{center}
\mbox{\psboxto(10cm;10cm){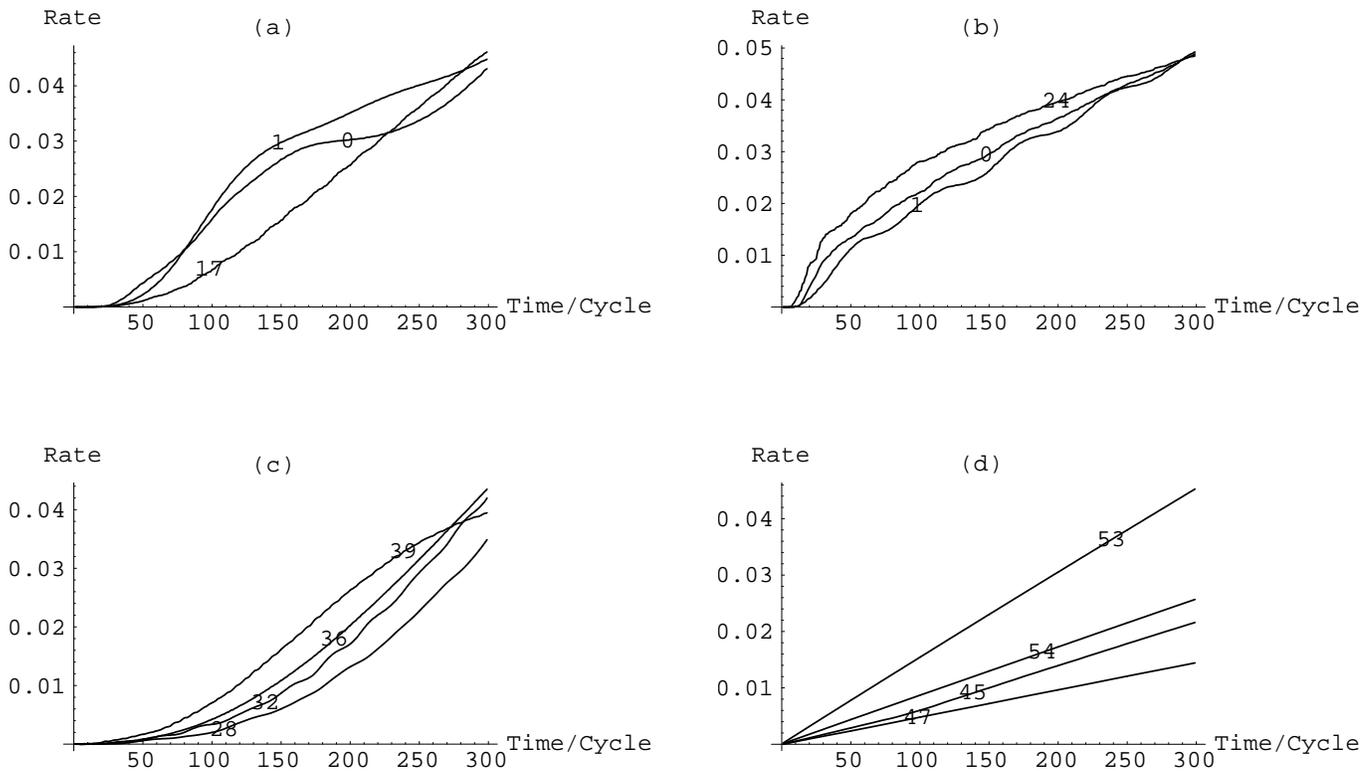}}
\caption[1]{Dissociation histories of four types observed at critical
amplitudes $A = A_c$ and  laser frequency $\Omega =
0.9$; (a)
the curves rise smoothly; (b) the curves rise sharply; (c) curves that are
distinct from those of neighbouring states; (d) straight lines.}
\end{center}
\end{figure}

The dissociation histories
from Eq. (\ref{P'}) 
were
calculated at $\Omega=0.9$. 
Fig.1 shows
dissociation
histories of four types at critical amplitudes,
i.e., the largest amplitude at which the molecule just fails to be dissociated
within 300 vibrations.
There are several such critical amplitudes for some low-lying initial states (
{\it vide infra} ).
This classification is not rigorous,
but represents at least some typical dissociation histories;
it results from investigation of more than 500 curves obtained for the
calculation for Fig.3. \cite{TYJ} shows other curves. A brief
description of each type follows.
For the first type the curve rises smoothly, whereas for the  second type 
the curve rises sharply. 
For those initial states with two transitions 
the smaller value corresponds to the first type
whereas 
the larger value of $A_c$ corresponds to the second type.
All those states belonging to the second type are expected to 
have a smaller value of $A_c$, although
only part of them were found.
The third type is essentially the same as the
first type but their neighbours belong to the second type. The fourth
type belongs to the initial states exceeding 44; the dissociation history is almost a
straight line.
Except for the latter type, for which the figure cannot display 
clearly, the reason for
a greater dissociation period is not only that 
the slope of the curve was altered, but
that there were several flat regions in which there was little molecular
dissociation.
As a further investigation for the third type, Fig.2 compared
at the same amplitude
with initial states near $n_0=28$, indicating that some initial states such
as
$n_0 = 28$ dissociate more readily than their neighbours. 
These states
correspond to the ionisation windows proposed by \cite{BS1}.
There are indications that some states dissociate more easily than
others. The dissociation process might be taken to have two steps; i.e., the
population first transferred to those easy states 
then
to other states.
Generally the populations deployed themselves soon after the
process began and stayed well localised, 
as \cite{WP} observed.
\cite{S,PP} discussed that an initial set-up time is required for any periodic system is a
general property, because the system has to sense 
the periodic perturbation.
Various quantities such as the 
inter-atomic distance of initial states and the average variation of energy
with time  were also computed.
A preliminary result on the inter-atomic distance shows that,
immediately after the field is switched on, states with $n_0 > 25$ move
in a direction distinct from that for $n_0 < 25$ away from the inter-atomic equilibrium position.

\begin{figure}
\begin{center}
\mbox{\psboxto(5cm;5cm){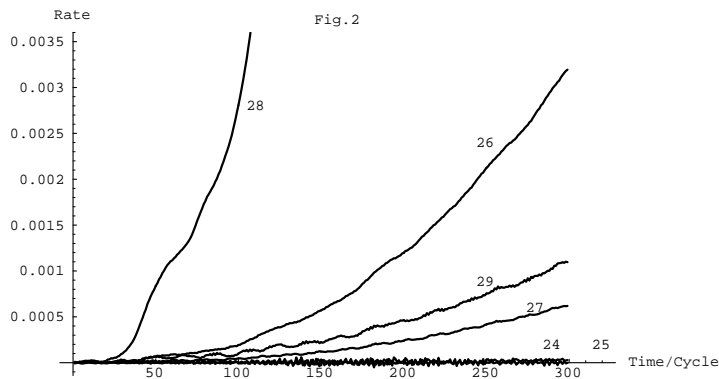}}
\caption[2]{Dissociation histories of initial states from 24 to 29 at A=0.01 and $\Omega =
0.9$; the state with $n_0$ = 28 dissociate more readily than its neighbouring states.}
\end{center}
\end{figure}

\begin{figure}
\begin{center}
\mbox{\psboxto(5cm;5cm){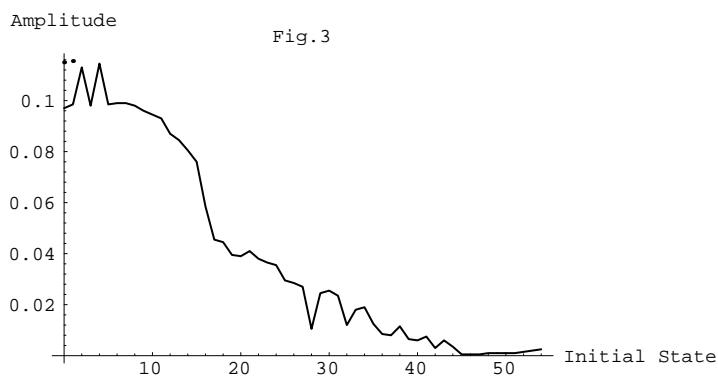}}
\caption[3]{Critical amplitude ($A_c$) for various initial states at $\Omega
=0.9$; 
the Hamiltonian is Eq. (\ref{H1}) and the dissociation rate is
Eq. (\ref{P'}); 
for low-lying states which have two values of $A_c$ only the
lower one is connected.}
\end{center}
\end{figure}

In Fig.3 the critical fields required for the molecule to become
dissociable
is plotted. 
For multi-critical states the lowest ones are connected and the others marked.
Therefore the plot provides only a sufficient criterion for dissociation, not a
necessary condition. The reason is  that as 
we use numerical methods
we can know only the points that we computed.
The amplitude resolution of our calculation is $\Delta A= 0.0005$. If there
were a narrow transition less than 0.0005 one might not be able to find it, 
because such a point would imply experimentally a laser of precise power.
We find at most two
critical amplitudes in the present case while in another study using two-color laser fields
we find up to three.
The multi-critical amplitude is also a result of our selection of the critical
dissociation period to be 300.
However, this point implies  that decreasing field
strength does not necessarily increasing the dissociation period.
\cite{LHVY}
noted that the finite size of $\hbar$ probably places a characteristic scale
in parameter
space less than which the dissociation boundary ceases to show fractal
structure. 
Another feature of the figure is that the largest value of $A_c$ is several orders of 
magnitude larger than the smallest one. 
A calculation 
shows that for $ n_0 \geq 6$ the photon
energy becomes larger than the energy difference between adjacent states.


\begin{figure}
\begin{center}
\mbox{\psboxto(5cm;5cm){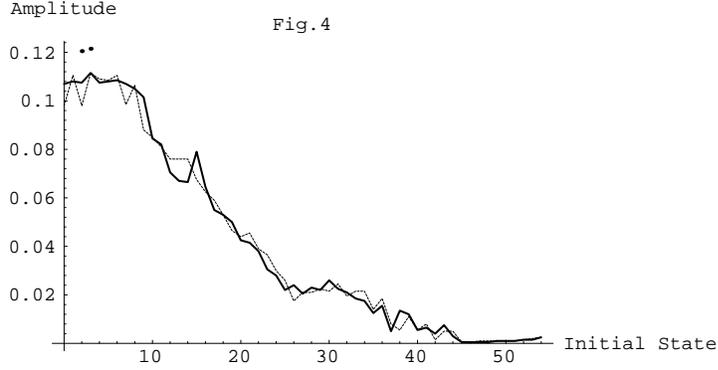}}
\caption[4]{Critical amplitude ($A_c$) for various initial states at (a) $\Omega
=0.89 $ (solid thick line) and (b) $\Omega =0.91$ (thin dashed line);
the Hamiltonian is Eq. (\ref{H1}) and the dissociation rate is
Eq. (\ref{P'}).}
\end{center}
\end{figure}

To discover whether $\Omega =0.9$ is the most efficient frequency for all
initial states we tested $\Omega=0.89$ and $\Omega=0.91$. 
The critical amplitudes are plotted in Fig.4. As we had insufficient 
computer time to test all states and all frequencies, we selected $n_0= 1, 17 $ and 24 
to find whether values of $A_c $ alter for frequencies from 0.8 to 1.1 at interval 0.01.
The plots appear in Fig.5.

\begin{figure}
\begin{center}
\mbox{\psboxto(5cm;5cm){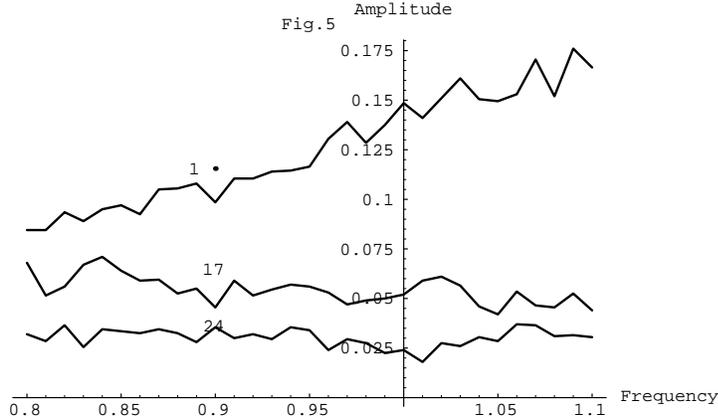}}
\caption[5]{Critical amplitude ($A_c$) vs. frequency of initial
states (a) $n_0 = $ 1, (b) $n_0 = $ 17 and (c) $n_0 = $ 24.}
\end{center}
\end{figure}


\subsection{Exponential coupling}
The perturbation in Eq. (\ref{H1}), although simple and easily done
by the Floquet method, has several problems in relation to reality. Firstly,
the molecular dipole and laser coupling in equation Eq. (\ref{H1})
assumes a infinite range of constant coupling, 
which is satisfactory for an atomic system
but not for a real molecule.
Secondly, as \cite{SS} has noticed, the field should be zero at $t=0$ and 
gradually increases from zero to a finite value; this problem is 
called adiabatic switching. 
Therefore, another exponential form
of coupling
\begin{equation}
H = H_0 -{{A \Omega} \over 2}{(z+a) e^{-(z+a)/b}} \sin (\Omega t) \sin^2
({{2 \pi t} \over T}),
\label{MH}
\end{equation}
was used to test whether there is any effect of the approximation.
The problem of the definition of dissociation rate was
pointed out above (below Eq.(\ref{P'})); here we alter it to comply with
\cite{HM2}
\begin{equation}
 P_{diss} (t) = 1 - < \psi(t)  | \psi(t) >.
\end{equation}
This definition is appropriate if coherent excitation becomes
important.
The variables $a$ and $b$ were chosen to have the values 2 and 1
respectively in the calculation.
\cite{HM2,HM1,TM2,TM1} have also used coupling functions of
similar form.
We take $T$, the pulse duration, to be 300 optical cycles. 
This choice has the
advantage that, when $A=A_c$, $T_d$ is almost 300 cycles and the field
strength is almost zero at that time. Hence there is little effect of
the definition of dissociation rate.
The resulting critical front appears in Fig.6 as a solid thick line.
There is no essential distinction from Fig.3 and Fig.4 
except a scaling parameter 
that can be
adjusted by choosing a and b. 
For lower 
initial states
there is more variation than for higher initial states. 
To assess whether $T$ is large enough, a longer period $T = 600$ is
tested. The result is plotted in Fig.6 with a thin dashed line; the two
lines show
similar behaviour.

\begin{figure}
\begin{center}
\mbox{\psboxto(5cm;5cm){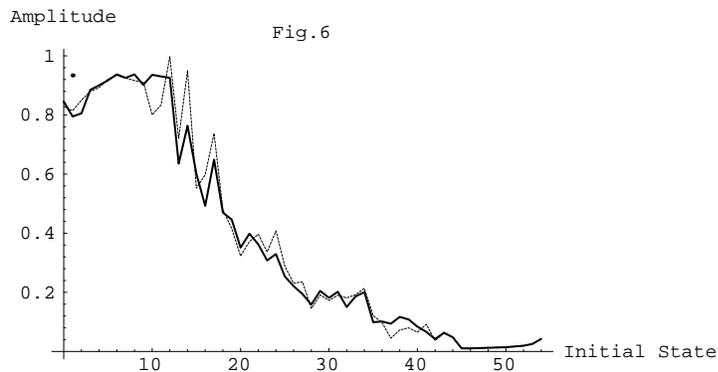}}
\caption[6]{Critical amplitude ($A_c$) of various initial states and
modified potential Eq. (\ref {MH}) at $\Omega
=0.9$; for the solid thick line $T=300$ whereas for the thin dashed line 
$T=600$.}
\end{center}
\end{figure}

\section{Summary}

In conclusion, thresholds of vibrational dissociation of NO are 
considered in this paper.  The main result of this paper is depicted in
Fig.6.
Two perturbations
and definitions of dissociation rate lead to similar curves with the
transition from high $A_c$ to low $A_c$ shifted slightly to higher initial
states.
Laser fields and couplings of 
other kinds were 
tested for which similar curves were obtained. 
Therefore, 
as our main concern is the critical amplitude which is
an indirect result,
we conclude that our results are
robust to variation of the potential function.
Variations of other types such as the laser
duration and definition of dissociation threshold are less significant. 
A rough estimate of the decrease of $A_c$ is given.
The assumption behind this model is that no other electronically nonadiabatic
effects will occur first.
One must take care in comparing  the results presented in this work with 
those from  experiment. For
instance, because of the finite cross section of the beam, not all molecules
experience the same field strength. The ignorance of rotational
influence should also be remembered. The additional selection
rule in the rotating case may make the progression to dissociation
harder.

\section{Acknowledgments}
 Finally I wish to express my gratitude to Drs. T.F. Jiang and J.F. Ogilvie
for discussions and Professor How-Sen Wong for
encouragement.

\end{document}